%% file: main.tex
\documentclass[10pt,conference]{IEEEtran}
\IEEEoverridecommandlockouts
\usepackage{cite}
\usepackage{amsmath,amssymb,amsfonts}
\usepackage{algorithmic}
\usepackage{graphicx}
\usepackage{textcomp}
\usepackage{url}
\usepackage{xcolor}
\usepackage{booktabs}
\usepackage{paralist}
\usepackage{dirtytalk}
\usepackage[caption=false]{subfig}
\usepackage{microtype}

\def\BibTeX{{\rm B\kern-.05em{\sc i\kern-.025em b}\kern-.08em
    T\kern-.1667em\lower.7ex\hbox{E}\kern-.125emX}}

\begin{document}

\title{Temporal Discounting in Software Engineering: \\ A Replication Study}

\author{
  \IEEEauthorblockN{Fabian Fagerholm}
  \IEEEauthorblockA{University of Helsinki, Finland \\
    Blekinge Inst. of Tech., Sweden \\
    Univ. of Toronto, Canada \\
    {fabian.fagerholm@helsinki.fi}}
\and
\IEEEauthorblockN{Christoph Becker}
\IEEEauthorblockA{Faculty of Information \\
  University of Toronto, Canada \\
  {christoph.becker@utoronto.ca}}
\and
\IEEEauthorblockN{Alexander Chatzigeorgiou}
\IEEEauthorblockA{Dept. of Applied Informatics \\
  Univ. of Macedonia, Greece \\
  {achat@uom.gr}}
\and
\IEEEauthorblockN{Stefanie Betz}
\IEEEauthorblockA{Furtwangen Univ., Germany\\
  {stefanie.betz@hs-furtwangen.de}}
\and
\IEEEauthorblockN{Leticia Duboc}
\IEEEauthorblockA{La Salle Univ., Spain \\
  {l.duboc@salle.url.edu}}
\and
\IEEEauthorblockN{Birgit Penzenstadler}
\IEEEauthorblockA{CSULB, Long Beach, USA  \\
LUT, Lappeenranta, Finland \\
  {birgit.penzenstadler@csulb.edu}}
\and
\IEEEauthorblockN{Rahul Mohanani}
\IEEEauthorblockA{Dept. of CSE \& HCD \\ IIIT Delhi, India \\ 
  {rahul.mohanani@iiitd.ac.in}}
\and
\IEEEauthorblockN{Colin C. Venters}
\IEEEauthorblockA{University of Huddersfield, UK\\
  {c.venters@hud.ac.uk}}
}

\IEEEpubid{\makebox[\columnwidth]{\mbox\hfill} \hspace{\columnsep}\makebox[\columnwidth]{ }}

\maketitle


\begin{abstract}

\textit{Background:} Many decisions made in Software Engineering practices are intertemporal choices: trade-offs in time between closer options with potential short-term benefit and future options with potential long-term benefit. However, how software professionals make intertemporal decisions is not well understood. 

\textit{Aim:} This paper investigates how shifting time frames influence preferences in software projects in relation to purposefully selected background factors. 

\textit{Method:} We investigate temporal discounting by replicating a questionnaire-based observational study. The replication uses a changed-population and -experimenter design to increase the internal and external validity of the original results.

\textit{Results:} The results of this study confirm the occurrence of temporal discounting in samples of both professional and student participants from different countries and demonstrate strong variance in discounting between study participants. We found that professional experience influenced discounting. Participants with broader professional experience exhibited less discounting than those with narrower experience.

\textit{Conclusions:} The results provide strong empirical support for the relevance and importance of temporal discounting in SE and the urgency of targeted interdisciplinary research to explore the underlying mechanisms and their theoretical and practical implications. The results suggest that technical debt management could be improved by increasing the breadth of experience available for critical decisions with long-term impact. In addition, the present study provides a methodological basis for replicating temporal discounting studies in software engineering.
\end{abstract}


\section{Introduction}
Software practitioners are people, but are often expected to behave differently and ``more rationally'' than average people: they are supposed to consider criteria more numerically than others, make estimates with quantified uncertainty, perform trade-off analysis using multiple weighted factors, and pursue the design decisions with the assumed ``highest value''. What they do in practice is another matter~\cite{ralph_two_2018, becker_trade-off_2018}. 

Many decisions made in Software Engineering practice are intertemporal: they involve trade-offs in time between closer options with potential short-term benefit and future options with potential long-term benefit. ``Temporal discounting'' is the degree to which an outcome's distance in time affects its perceived value. This is most explicitly visible in technical debt management, but is by no means restricted to that area. For example, many architectural trade-offs manifest at different timescales; the prioritization of features when planning iterations implies trade-offs in time; and similarly, refactoring decisions, the development of test suites, and code documentation choices all involve costs and benefits occurring at different time frames. Most explicitly, the source of technical debt has been located in ``decisions that are `expedient' in the short term but \ldots{} costly in the long term'' \cite{mcconnell2007, cunningham1992}, implying the temporal nature of those decisions and the tendency for temporal discounting of future options.

The complexity of factors that influence people's behavior in SE requires special care when we study how people make decisions. For instance, the decision-making process in software project management is largely based on human relations~\cite{dacunha2016}. Software project management decision-making can be characterized as knowledge sharing and is participatory, with the project manager acting as a facilitator. Effective SE leaders delegate decision-making and act to shape and cultivate effective decision-making behavior -- empowering software developers to make autonomous and informed decisions~\cite{kalliamvakou2019}.

This complex and distributed nature of SE project decision-making means that decision-making support must address not only project managers, but several other roles involved in decision-making. Interactions between the decision-making environment and psychological factors at the individual level may cause omission behavior, where developers forego established practice and methods, and choose short-term, quick, ad-hoc solutions with impact on product quality and other adverse effects in the long term~\cite{ghanbari2018}. One reason may be that the methods are only suitable for some of the situations that arise in real software projects. The majority of studies on technical debt in software engineering have focused on prescriptive approaches. Little focus has been placed on how software developers actually make such decisions \cite{becker2019}.

How software professionals really make intertemporal decisions is not well understood, but research from the field of Judgment and Decision Making (JDM) provides strong methods for exploring this question \cite{keren_wiley-blackwell_2015}. While the methods and theories of JDM have not been applied yet to examine intertemporal choices in SE in depth, a previous study investigating intertemporal choice in technical debt management found evidence for temporal discounting: Developers' valuation of probable future outcomes was significantly reduced when the time frame was extended~\cite{becker2019}. The study provided initial evidence to answer the questions: \say{How do software practitioners discount uncertain future outcomes?} and \say{How strong are temporal discounting effects, and how much do they vary?}.

These initial results require robust replication to assess their validity and raise many questions that stimulate further investigation to ascertain what factors may influence temporal discounting among software practitioners. Catering to the needs of different stakeholders in SE requires strong empirical evidence~\cite{kitchenham2004}, and replication is a central part of accumulating reliable evidence~\cite{dyba2005}. Robust SE theories with broad applicability and with consideration of psychological insights calls for rigorous research and particularly strong empirical grounding through original studies and their replication~\cite{shepperd2018}. Understanding the real-life context in SE is crucial for building generally useful theories and results~\cite{petersen2009}, and that contextual understanding can be improved through a human factors perspective~\cite{lenberg2015}. We believe it is important to establish some fundamental results before expanding into more complex perspectives. For these reasons, this study aims for improved rigor by replicating the original study.

The current study pursues three main research questions.
\begin{enumerate}
    \item [RQ1:] Is the discounting effect found in the original study confirmed in other samples? 
    \item [RQ2:] Are there differences between samples from different countries? 
    \item [RQ3:] Do factors such as professional education, professional experience and team agility play a role in temporal discounting?
\end{enumerate}

We replicated the original study to investigate whether temporal discounting occurs in a software project management task and whether purposefully selected background factors influence discounting. The replication uses a changed-populations and -experimenter design to increase internal and external validity.

The replication results confirm the occurrence of temporal discounting among software practitioners and students. They also demonstrate strong variance in discounting between study participants, as found in the original study. The replication contributes new information regarding the influence of background factors on temporal discounting: age, prior education and workplace training, professional experience, and perceived agility of the development team. Interestingly, the only factor among those investigated with significant influence on discounting is the breadth of professional experience. Finally, the replication demonstrates the occurrence of discounting in samples of both professional and student participants from different countries and thus cultural backgrounds.

This replication study contributes to the field by introducing methods and theories from JDM research with central relevance to SE. The results provide strong empirical support for the relevance and importance of temporal discounting in SE and the urgency of targeted interdisciplinary research to explore the underlying mechanisms as well as theoretical and practical implications. The present study thus provides a methodological basis for replicating temporal discounting studies in SE.







\section{Background}

Intertemporal choice research has been conducted in multiple fields, but a previous study~\cite{becker2019} is among the first to examine the topic in SE. In this section, we introduce the concepts of intertemporal choice and temporal discounting, discuss the results of a study examining temporal discounting in an area of SE, and provide background on replication studies.

\subsection{Intertemporal choice}

Intertemporal choices -- ``decisions involving trade-offs among costs and benefits occurring at different times''~\cite{frederick2002} -- abound in software development. Perhaps the most obvious scenarios occur in technical debt management, where the trade-offs appear both explicitly and implicitly. Many TD management decisions must explicitly and directly address the temporal trade-off. Temporal discounting also plays an implicit role in behaviours where temporally distant outcomes are disregarded without a conscious choice, which is often considered as a source of TD \cite{mcconnell2007}.

Intertemporal choice is a very active research area in psychology, behavioural economics, neuro-economics, management science, marketing, and other fields \cite{frederick2002, soman2005, loewenstein2003, loewenstein2008}. Intertemporal choice research takes neurological, psychological and sociological perspectives to examine and explain how individuals and groups makes choices with intertemporal aspects; to understand the conditions under which humans disproportionately discount the future; to model and predict consumer and other human behaviour; or to understand how the architecture of choice influences outcomes. The precise mechanisms that influence temporal discounting are still not completely understood, and the best theoretical models to describe and explain it remain disputed, despite decades of research~\cite{frederick2002}. Nevertheless, the body of research on intertemporal choice provides powerful theories, research methods, experiment designs, empirical guidelines, and conceptual frameworks that support scientific exploration, understanding, modelling, and predicting how people will make intertemporal choices \cite{loewenstein2003, loewenstein2008, soman2005, weber2006}.

In JDM, the concept of a ``decision'' is much broader than the narrow conception of selecting among a set of options based on explicitly defined criteria. In fact, that is often not how people make decisions~\cite{lipshitz_taking_2001,isenberg_how_1984, montgomery_how_2005,zannier_model_2007,becker2017,zsambok_naturalistic_1997}. Instead, decision making involves such aspects as generating options, mentally simulating outcomes, and devising courses of action~\cite{klein1998,klein_naturalistic_2008,lipshitz_taking_2001}. The methods available to examine these cognitive and social processes cover a wide spectrum ranging from experimental and quasi-experimental methods, suited to empirically establishing phenomena and filtering plausible explanations, to Cognitive Task Analysis (CTA) studies with rich accounts of the macro-cognitive systems that give rise to real-world behavior~\cite{crandall_working_2006}. This implies that despite the valid critique of the limitations of the narrow view of decision making as a lens for understanding systems design and Software Engineering practice \cite{dorst_creativity_2001,ralph_characteristics_2016}, the conceptual frameworks of JDM and CTA retain immense relevance to the empirical study of SE practice \cite{ralph_two_2018}.

\subsection{Temporal discounting}

A variety of methods have been used in empirical studies to examine intertemporal choice~\cite{frederick2002}. Hundreds of studies have explored the intertemporal choice behaviour \emph{of consumers} in particular~\cite{frederick2002}. A typical and frequently used basic experiment examines an explicit trade-off between a monetary reward at one point in time, and a (higher) monetary reward at a later point in time. Such experiments establish a \emph{discount rate} that reflects how much higher a later reward must be to be considered equally valuable as a closer reward. The discount rate describes, in numerical terms, how experiment participants discount future outcomes. Real-world temporal discounting behaviour has been effectively predicted in laboratory experiments in many domains, including ``credit card debt, smoking, exercise, body-mass index, and infidelity''~\cite[p.\,3]{hardisty2011}.

Different studies have found an extreme range of discount rates, ranging from negative to $\infty$, with most results ranging from 0\% to 500\%~\cite{frederick2002}, partly due to differences in measurement methods~\cite{frederick2002, hardisty2011}. While some individuals in some situations may genuinely exhibit a negative discount rate, the expected behaviour generally involves a positive rate, perhaps since the present is more salient than the distant future. Many factors can influence a person's temporal discounting, including several anchoring and priming effects, the framing of outcomes as losses or gains, the order in which choices are presented, the viscerality of outcomes (how vividly the person can imagine the outcome), and psychological distance~\cite{hardisty2011, weber2006}. On its own, however, temporal discounting does not explain \emph{why} the timing of outcomes is associated with how individuals value them, it only demonstrates that such discounting occurs.

How to elicit the discount rate is of central importance to experimental studies on temporal discounting \cite{frederick2002}. Two common approaches are choice and matching tasks. A \emph{choice task} means that participants choose the preferred outcome among two options. A \emph{matching task} entails ``filling in a blank'' to indicate a quantity, e.g., a monetary reward, that would make one outcome equivalently attractive for the participant to another outcome at a different point in time. Choice tasks are often presented in a sequence with varied outcome parameters to narrow upper and lower bounds to allow computing an indifference point~\cite{hardisty2011}. Matching tasks directly ask for the indifference point. Indifference points for different time horizons form the basis for computing the discount rate.

Choosing the presentation of the task or question that acts as the stimuli for a temporal discounting experiment, and the variables that are manipulated within and across participants, are central research design choices. Another important consideration is how to calculate implied discount rates from observed behaviours. Traditionally, intertemporal choice research has focused on a simple normative model of \emph{Discounted Utility} (DU) developed by Samuelson~\cite{samuelson1937}. DU models the discounting process as exponential: time-consistent and with a constant discount rate. This is akin to the interest rate on loans and investments, which a ``rational'' decision-maker could supposedly use as a reference. The exponential model, while relatively simple, often does not fit empirical observations; numerous studies have demonstrated deviations from it~\cite{frederick2002}. This has led to the proposal of several other models of temporal discounting. Each model supplies a different way to calculate the discount factor from the observed indifference points. Still, they are all based on the amount that a person would require to prefer the future option (future value, $FV$), the amount available earlier, often immediately (present value, $PV$), and the time between the two options $t$. In the DU model, the annualized continuously compounded discount rate $DR_c$~\cite{samuelson1937} relies on the future value $FV$ defined as (\ref{eq:FV}).

\begin{equation}
  FV=PV \times e^{DR_c \times t}
  \label{eq:FV}
\end{equation}
Which can be solved as (\ref{eq:DR}) to obtain $DR_c$:
\begin{equation}
  DR_c(FV, PV, t) = \frac {\ln{\frac{F}{P}}} {t}
  \label{eq:DR}
\end{equation}

An alternative model, \emph{hyperbolic discounting}~\cite{mazur1987}, does not assume a constant discount rate~\cite{doyle2012}. Instead, assigned value falls rapidly for earlier delays, but more slowly for longer delays, depending on the parameters used. 

Different models yield different results and are based on different conceptions of the underlying discounting process. For example, some models exaggerate discounting differences between small time intervals, while others are less sensitive to differences. The choice of model is a research topic in its own right. Several papers (e.g.,~\cite{hardisty2011, samuelson1937, mazur1987, myerson2001}) discuss the merits of different models, propose new models or variants of existing models, and examine their fit to different sets of empirical data. Ultimately, the intertemporal choice task should be chosen based on the real-life phenomenon of interest~\cite{hardisty2011}, and the discount rate model must be chosen based on the data obtained while taking into account theoretical assumptions and comparability with related studies.

\begin{figure}
  \centering
  \includegraphics[width=\columnwidth]{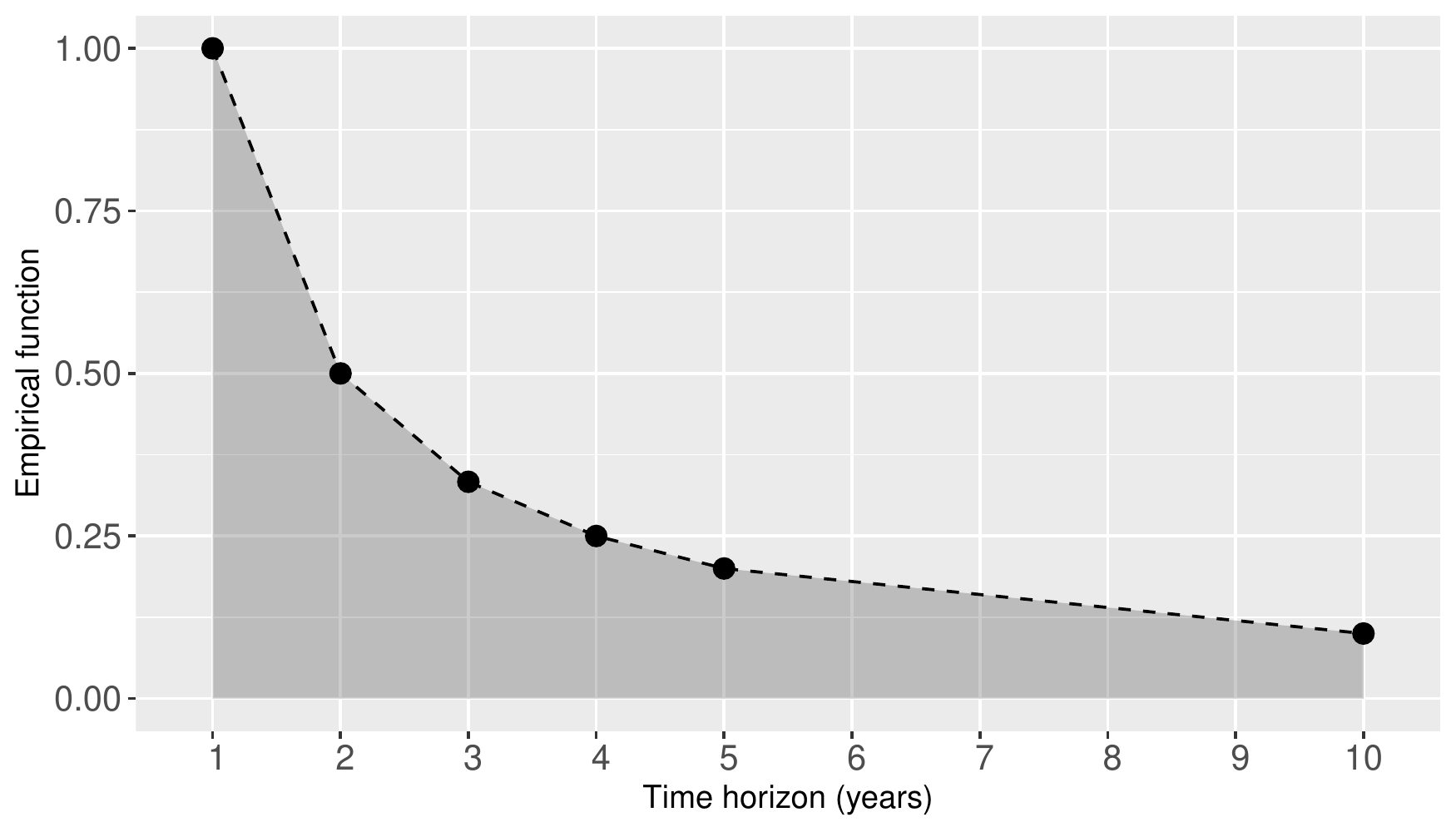}
  \caption{Example of the area under curve (AUC) approach for a single participant in this study. The curve shows that they value future options lower than present value; the value of the 10-year option is less than 20\% of the first. The total AUC for this participant, depicted by the shaded area under the curve, is $2.43$.}
  \label{fig:auc-example}
  \vspace{-0.2cm}
\end{figure}

Another approach to calculating the discount rate is to use the area under curve (AUC) approach~\cite{myerson2001}, which avoids many theoretical questions by not attempting to fit a curve to the data. Instead, it just summarizes the shape of the empirically observed subjective valuation. In this approach, time delays can be normalized as a fraction of the maximum delay, and the subjective values observed by the nominal amount (i.e., PV). An example is shown in Fig.~\ref{fig:auc-example}. The normalized values form trapezoid curve segments. The area of each segment is the discount rate for that time horizon, and is calculated as shown in (\ref{eq:AUC}) ($x_n$: different time horizons; $y_n$: observed values). The AUC approach may also be used to provide a single, overall discount rate by summing the areas of all segments. This allows comparison between participants and statistical testing of overall discounting across background variables.

\begin{equation}
  DR_{AUC} = (x_2 - x_1) \times \frac{y_1 + y_2}{2}
  \label{eq:AUC}
\end{equation}

\subsection{Previous study}

In December 2018, the first study \cite{becker2019} (hereinafter, the ``original study'') on temporal discounting in software engineering was performed. The goal was to investigate -- as in the current study -- how software practitioners discount uncertain future outcomes and whether they exhibit temporal discounting. An online questionnaire was administered to software developers from two large companies  (with more than 100 employees) in Greece. The responses allowed the extraction of the discount rates for 33 participants, with a mean age of 34.3 years and approximately 7 years of work experience on average. In the scenario that was presented, exactly as in the present study, two options were available on how to spend an upcoming week: the short-term option was to implement now a feature from the next iteration, while the long-term option was to integrate a new library with no instant benefit, but with a 60\% chance of saving future effort (see Fig.~\ref{fig:survey_task}). The matching task presented to the participants asked them to indicate the minimum amount of potential time saving (in person-days) they would require to choose the long-term option over the short-term one.

\begin{figure*}
  \centering
  \subfloat[Company 1.]{\includegraphics[width=0.5\textwidth]{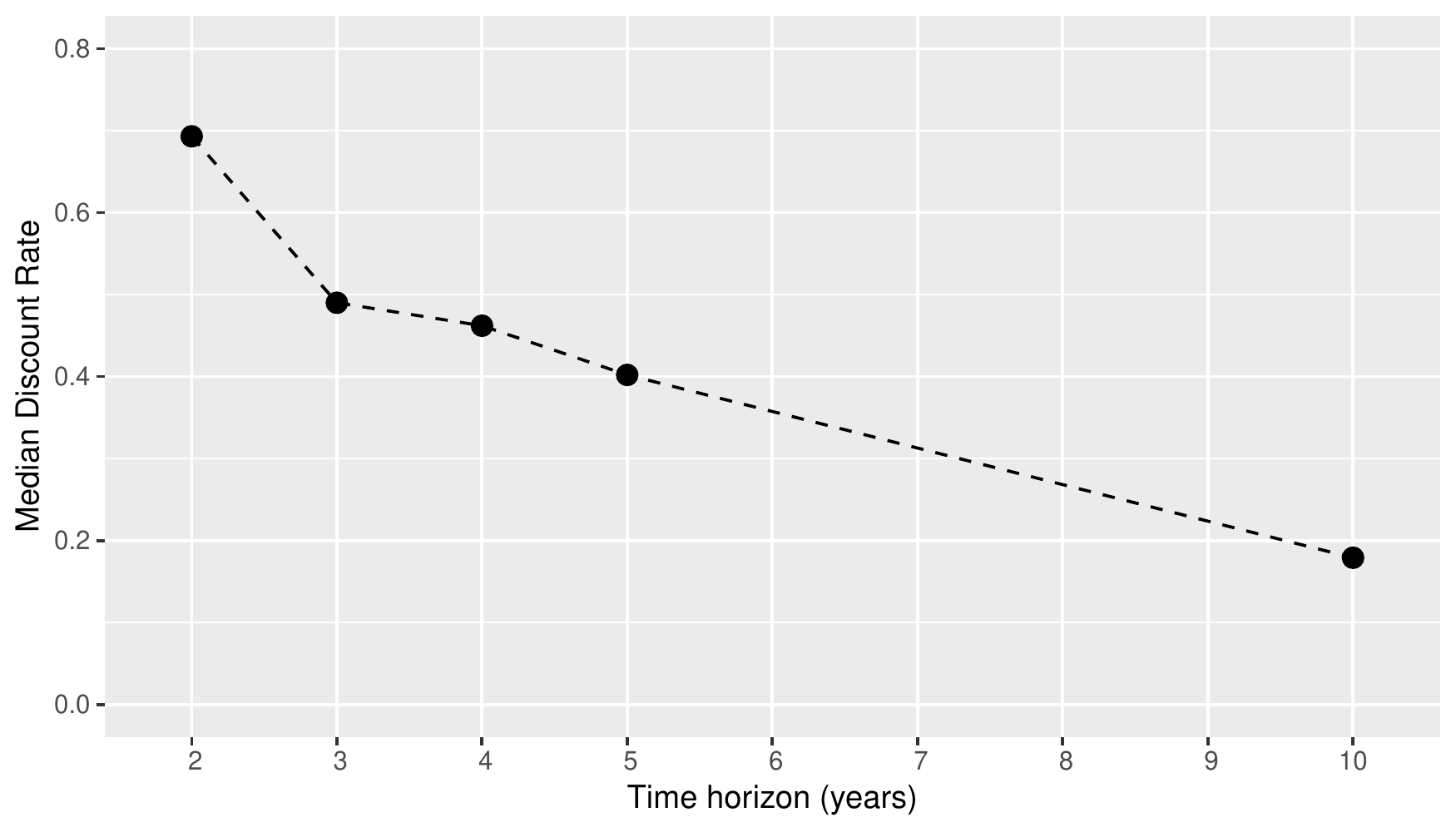}\label{discountratescomp1all}}
  \hfill
  \subfloat[Company 2.]{\includegraphics[width=0.5\textwidth]{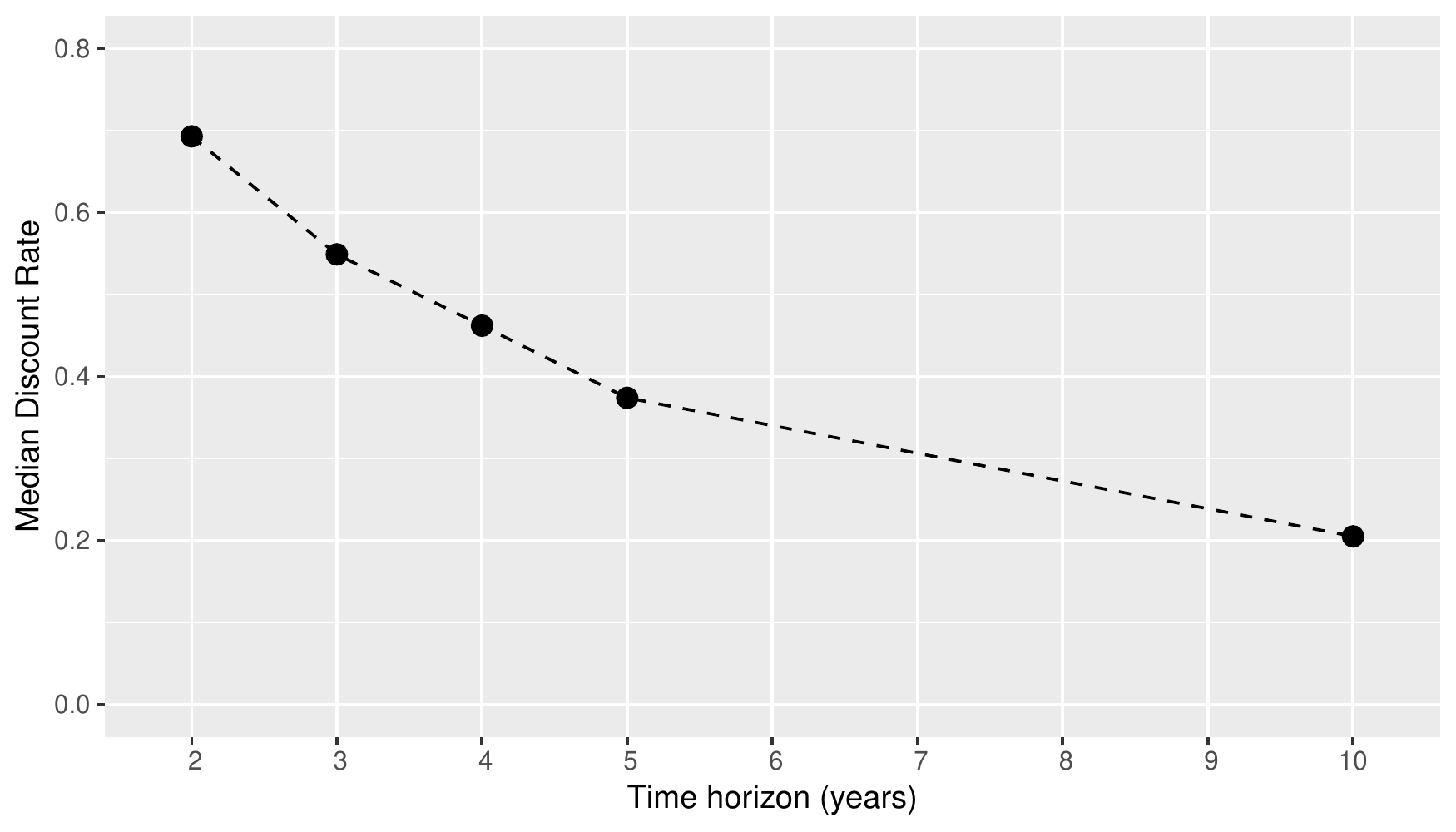}\label{discountratescomp2all}}
  \caption{Median discount rate as a function of time horizon (original study).}
  \label{fig:originalStudy}
  \vspace{-0.2cm}
\end{figure*}

The median discount rate for the employees of both companies, obtained using the exponential model, is shown in Fig.~\ref{fig:originalStudy}. Discounting is pronounced, and most pronounced for early time differences. For example, shifting the outcome from one year to two years involves a $70\%$ discount in both samples, while shifting it from four to five years only causes a roughly $40\%$ decrease in its implied value. Both rates are certainly much higher than common financial interest rates. The declining discount rate implies prevalent temporal discounting and is consistent with similar studies that analyzed consumer behavior in psychology and behavioral economics \cite{frederick2002,soman2005}. The analysis of participants' responses further revealed that for shorter time horizons, individual behaviors with respect to valuing uncertain future outcomes vary strongly, but for longer time horizons, they converge.

The results of the original study established the relevance of intertemporal choice theory and research for Software Engineering. At the same time, it raised a multitude of open research questions, especially with respect to the causal factors that potentially influence temporal discounting. We believe that systematically investigating these questions can drive the effective design and presentation of intertemporal choices in everyday Software Engineering situations.

For example, an interesting finding was the almost perfect match between the discount curves for the two companies, possibly implying that developers in a similar context and with analogous background value temporally discount outcomes in the same manner. Moreover, some participants did not exhibit any temporal discounting at all. These observations motivated us to replicate the original study to investigate further whether factors related to education, responsibilities, and length and breadth of experience influence temporal discounting. 

\subsection{Replication in Software Engineering}

Replication has been called a ``cornerstone of science'' from the perspective of researchers in many scientific fields~\cite{simons2014, shepperd2018}. Replication of experiments means repeating an experiment to validate its results and gradually build confidence in them~\cite{juristo2009}. Despite the general idea of replication being easy to understand, its meaning in SE research is not straightforward~\cite{carver2014}. In SE, replications have much in common with the social and behavioural sciences: close replications with nearly identical conditions are often not possible~\cite{miller2005, juristo2009}. Nevertheless, replication is an important part of validating findings and strengthening the research in empirical SE~\cite{carver2014}.

G\'omez et al.~\cite{gomez2014} describe two opposing views on replication in SE:
\begin{inparaenum}[1)]
\item that replications should retain only the hypothesis, and
\item that replications can retain more or less of the original.
\end{inparaenum}
They put forward a way of conceptualizing replications in more detail and classify them into \emph{literal}, \emph{operational}, and \emph{conceptual}. In the first type, the aim is to follow the original experiment as exactly as possible, and the replication is run by the same experimenters. The only differing aspect is that the sample, drawn from the original population, is different. This type of replication can serve to reduce sampling bias.

In an operational replication, four different dimensions may be varied~\cite{gomez2014}.
\begin{inparaenum}[1)]
\item Elements of the protocol may be varied to verify that the observed results are reproduced, thus addressing some specific biases.
\item The operationalization of cause and effect constructs may be varied to verify the bounds within which the results hold.
\item The population may be varied to verify the limits of the populations used in the original experiment.
\item The experimenter may be varied to verify their influence on results.
\end{inparaenum}

Finally, in a conceptual replication, a new protocol and new operationalizations are used by different experimenters to verify the original results. This type of replication can address several sources of bias, but may not identify what aspect of the original design may have introduced a bias, since more than one element is changed.

Replication is a fundamental part of software engineering research, but comes with several caveats. In internal replications (i.e., where one or more of the original authors take part) confirmatory results are alarmingly common, possibly due to researcher bias or inexact replication stemming from incomplete reporting~\cite{shepperd2018}. Combined with a great variety between replications even in the same, limited domains, this means that effect sizes and confidence limits are often not possible to determine. Shepperd et al.~\cite{shepperd2018} urge authors to more carefully document their replication designs, or to consider meta-analysis instead.

We note that meta-analysis is only possible when there is a substantial amount of existing research on a subject. Temporal discounting in software engineering has not been extensively studied before~\cite{becker2017}. Our chosen strategy is therefore to gradually expand the body of knowledge regarding temporal discounting by first replicating the original observational study design. We aim to lay the foundations for future studies that can expand the breadth and depth of the research on this topic, ultimately leading to a  solid research framework to study intertemporal decision-making in software engineering. The present paper is one step in this direction.

\section{Research design and analysis}

We replicated the original study \cite{becker2019} as an operational replication \cite{gomez2014} where we changed the population and partly varied researchers (changed-populations / -experimenters). We used the original study protocol, altered only in terms of collected background information. This replication design addresses internal and external validity threats of the original study and adds new information on some factors potentially influencing temporal discounting.

This study constitutes a replication but is not an experiment in the sense of a randomized controlled trial, as there is no treatment variation. Rather, it is an observational study that attempts to determine the existence of an effect (temporal discounting), and explore how selected background variables influence the effect. The theorized variations stem from individual differences and differences in respondents' environments.

\subsection{Questionnaire design}

The questionnaire from the original study \cite{becker2019} was used with some modifications to the background section. Participants saw a scenario description (see Fig.~\ref{fig:survey_task}) with two options:
\begin{inparaenum}[1)]
\item spend software project time earlier on implementing a planned feature (a short-term option); or
\item integrate a software library with potential long-term benefit in terms of reduced maintenance effort.
\end{inparaenum}
The scenario constituted a matching task in which they indicated the minimum potential time-saving they would require to choose the long-term option over the short-term option. The former was specified as having a 60\% chance of being realized to avoid additional discounting due to the lack of precise uncertainty \cite{frederick2002}. In the questionnaire, the scenario was first presented as a 1-year project to establish a baseline preference (present value, PV) free of priming from the consideration of different time-frames. The scenario was then presented again with a varying project time-frame of 1, 2, 3, 4, 5, and 10 years. The answer from the 1-year time-frame from this second presentation was not used in the analysis.

\begin{figure}
    \centering
    \fbox{\begin{minipage}{0.97\columnwidth}
    \centering
    \fbox{\begin{minipage}{0.97\columnwidth}
\footnotesize\sffamily
You are managing an \underline{N-years} project. You are ahead of schedule in the current iteration. You have to decide between two options on how to spend your upcoming week. Fill in the blank to indicate the least amount of time that would make you prefer Option 2 over Option 1.

\mbox{}

Option 1: Implement a feature that is in the project backlog, scheduled for the next iteration.
(five person days of effort).

\vspace{1mm}

Option 2: Integrate a new library (five person days effort) that adds no new functionality but has a 60\% chance of saving you \underline{\hspace{1em}} person days of effort over the duration of the project (with a 40\% chance that the library will not result in those savings).
\end{minipage}}

\vspace{1mm}

\begin{minipage}{0.99\columnwidth}
\footnotesize\sffamily
(The only difference here is the timeframe.)
\end{minipage}

\vspace{2mm}

\begin{minipage}{0.99\columnwidth}
\footnotesize\sffamily

For a project time frame of 1 year, what is the smallest number of days that would make you prefer Option 2?
\underline{\hspace{2ex}}
\vspace{2mm}

For a project time frame of 2 years, what is the smallest number of days that would make you prefer Option 2?
\underline{\hspace{2ex}}
\vspace{2mm}

For a project time frame of 3 years, what is the smallest number of days that would make you prefer Option 2?
\underline{\hspace{2ex}}
\vspace{2mm}

For a project time frame of 4 years, what is the smallest number of days that would make you prefer Option 2?
\underline{\hspace{2ex}}
\vspace{2mm}

For a project time frame of 5 years, what is the smallest number of days that would make you prefer Option 2?
\underline{\hspace{2ex}}
\vspace{2mm}

For a project time frame of 10 years, what is the smallest number of days that would make you prefer Option 2?
\underline{\hspace{2ex}}
\end{minipage}
\end{minipage}}
    
    \caption{Intertemporal choice task questionnaire (excerpt).}
\vspace{-0.3cm}
    \label{fig:survey_task}
\end{figure}

The data allowed us to validate the finding of temporal discounting in the original study. The demographic section had a few differences compared to the original. In addition to gender and age, we asked more detailed questions about education (whether academic or professional development), professional experience, perceived agility in the respondent's team, and work experience. These came after the intertemporal choice scenario and should not affect the main results on temporal discounting. The demographic section is summarized in Table~\ref{tab:survey-demographics} and the full questionnaire is given in an on-line replication package~\cite{fagerholm2019dataset}.

\subsection{Replication design}

The replications were set up as data collection rounds conducted by different researchers in different populations. Five researchers deployed the replication in 16 different population sets, 12 of which were companies, 2 professionals from different companies in two different countries, and 2 student populations (see Table~\ref{tab:samples}). One researcher was involved in the original study, while the four others were not. Each replication used a separate on-line questionnaire, identical except for the introduction text, which was slightly customized with contact information to the replicating researcher. The questionnaires for students had a slightly different instruction wording in the demographic section to better reflect the relationship with work that students may have -- some may have worked in the software industry while others may not.

\begin{table}
  \centering
  \caption{Summary of demographic questionnaire items.}
  \label{tab:survey-demographics}
  \include{tables/survey-demographics}
\end{table}

The replication design allowed comparison of respondents from different company samples in different countries. However, we assumed that differences in participation rates would yield different sample sizes. For this reason, the design assumes that data from different samples are combined into purposefully constructed sets that can be compared, but that the entire set would be analyzed for the main questions of the study.

\begin{table}[b]
  \centering
  \caption{Participant sets and their sizes.}
  \label{tab:samples}
  \include{tables/descriptive-samplesizes}
\end{table}

\begin{figure*}
  \centering
  \subfloat[Responsibility]{\includegraphics[width=0.33\textwidth]{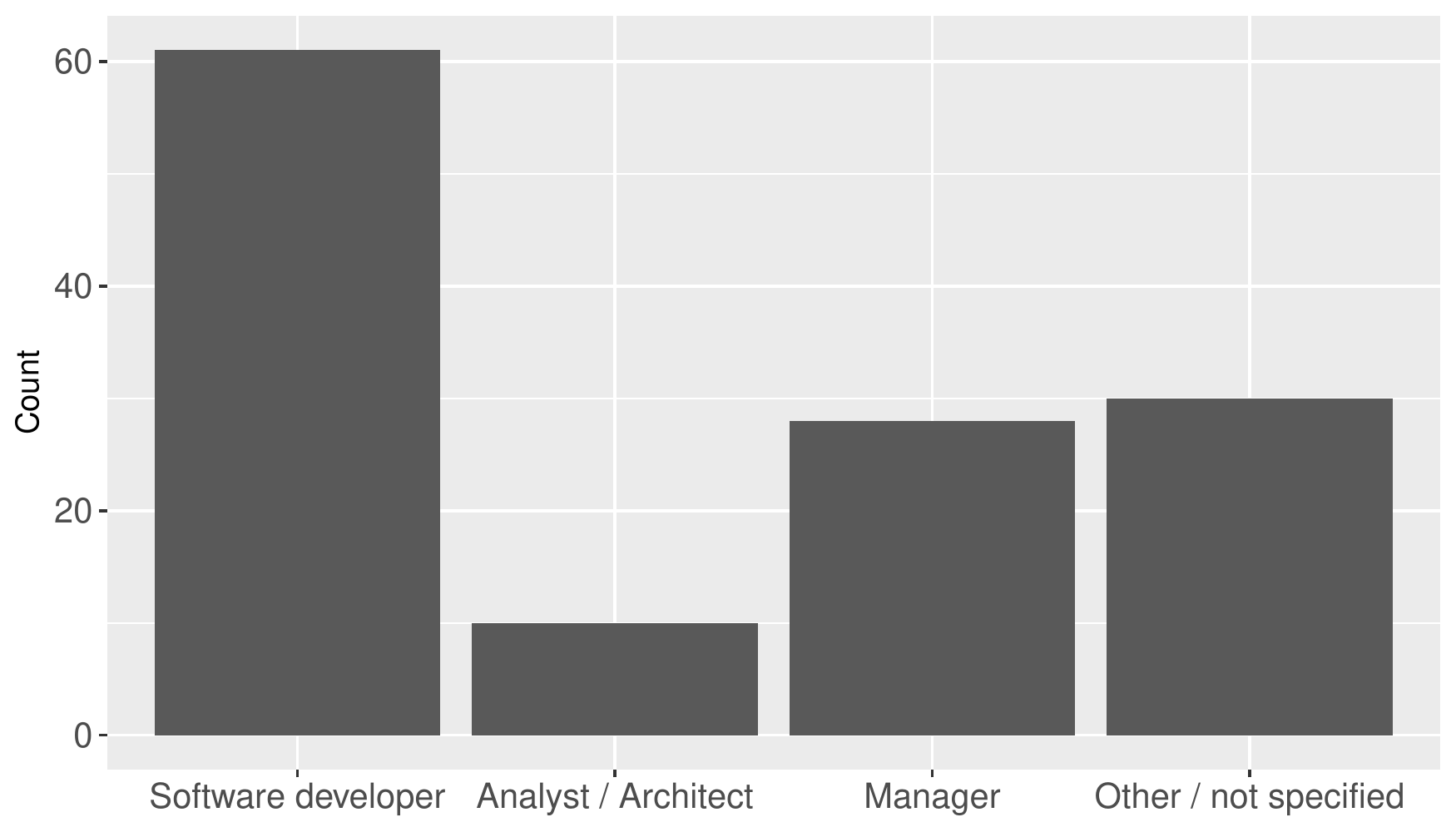}\label{roles}}
  \hfill
  \subfloat[Educational Qualification]{\includegraphics[width=0.33\textwidth]{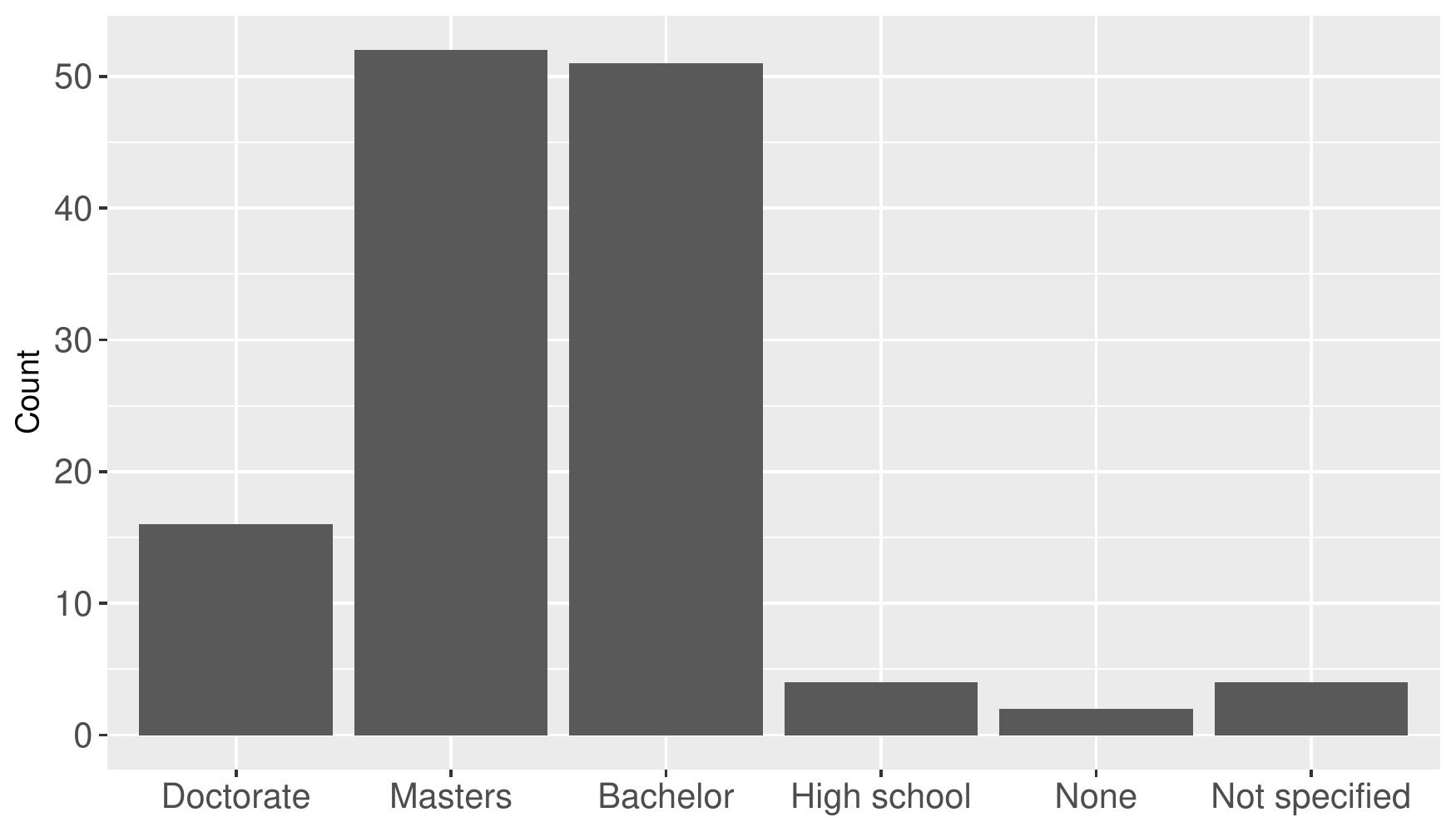}\label{education}}
  \hfill
  \subfloat[Work Experience]{\includegraphics[width=0.33\textwidth]{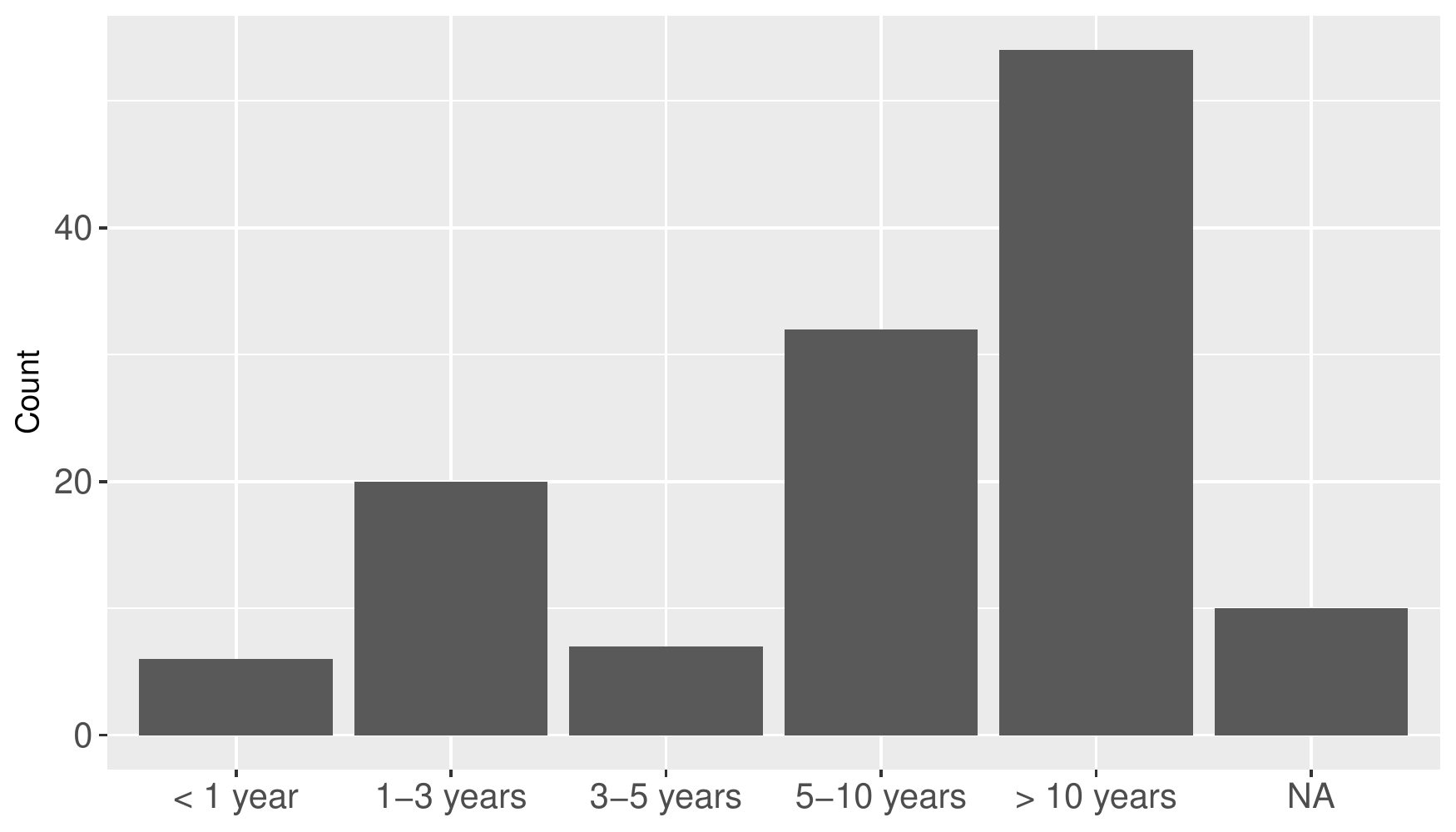}\label{workexperience}}
  \caption{Participant demographics.}
  \label{fig:demographics}
\end{figure*}

Because the data is partly collected by researchers not involved in the original study, we can to some extent decrease the potential researcher bias inherent in the original study. We established a clear protocol for the replications. Each replicating researcher was asked to describe the target sets of participants they would invite. One of the original researchers then created the online forms and customized their introduction and end texts in collaboration with each replicating researcher. The same researcher supported the replicating researchers throughout their data collection runs over email. Of the original researchers, only one collected data for the replication, and that researcher had no contact with the replicating researchers during the replication run. In this way, we attempted to both ensure that the interpretation of the study and replication protocols were correctly followed, but that involvement in data collection would not bias the results. This would not completely eliminate researcher bias, which would require that the study is replicated completely by different researchers.

\subsection{Data analysis}

We examined the questionnaire data using statistical methods as well as qualitative analysis for open text answers. We used simple content analysis of the short open text answers on company responsibility to broadly categorize respondents into comparable roles. We calculated the discount rate as a function of time horizons using the exponential model with annualized continuous compounding according to (\ref{eq:DR}). We calculated the overall discount rate using the area under curve for the empirical function, as shown in (\ref{eq:AUC}). Descriptive statistics, e.g., frequency and median, were used to examine the demographic data and describe the sample. Boxplots were used to gain an overview of the distribution of time-savings required by participants to prefer the long-term option. The median discount rate was plotted against the time horizon options to demonstrate the overall tendency. Individual discount rates were also plotted against the time horizon to examine individual differences. We examined the whole data set as well as selected subsets separately.

To examine the association between background variables and temporal discounting, we used the Kruskal-Wallis rank sum test to examine whether the overall discount rate, in terms of AUC, differed between different subsets by selected background variables. We employed the Anderson-Darling and Kolmogorov-Smirnov tests to examine similarity in shape between the AUC distribution of the different subsets, to ascertain whether interpreting the Kruskal-Wallis test as a difference in medians was appropriate. For continuous background variables, we used Pearson's correlation test to examine association with AUC, and for variables on a rank scale, we used Kendall's rank correlation test.

The choice of the exponential model was based on its use in the original study and the lack of evidence for model choice in the field. The model is commonly used in the intertemporal choice literature~\cite{hardisty2011}, is easy to calulate and replicate, and suffices to determine whether discounting occurs or not. The choice of AUC was based on its theory-neutrality~\cite{myerson2001}, which is a desirable characteristic in the light of the lack of evidence for model choice; the fact that it provides a comparable measure of the total discounting displayed in a participant's data; and its ease of calculation and replication.  The full data set and analysis scripts are available in a replication package ~\cite{fagerholm2019dataset}.

\section{Results}

We invited professionals and students in fields related to software development to participate in the study, either directly or through a company contact person. Twelve such sets of participants from six countries were invited during March to April 2019. Table~\ref{tab:samples} shows the number of participants in each set. We obtained a total of 129 usable responses. Two participants did not enter a number for the 1-year scenario in the first task, and we substituted those with responses from the second task. Some participants did not provide full background. We therefore either report missing responses as a separate category or use only complete responses as applicable.

\begin{figure}[b]
  \centering
  \includegraphics[width=\columnwidth]{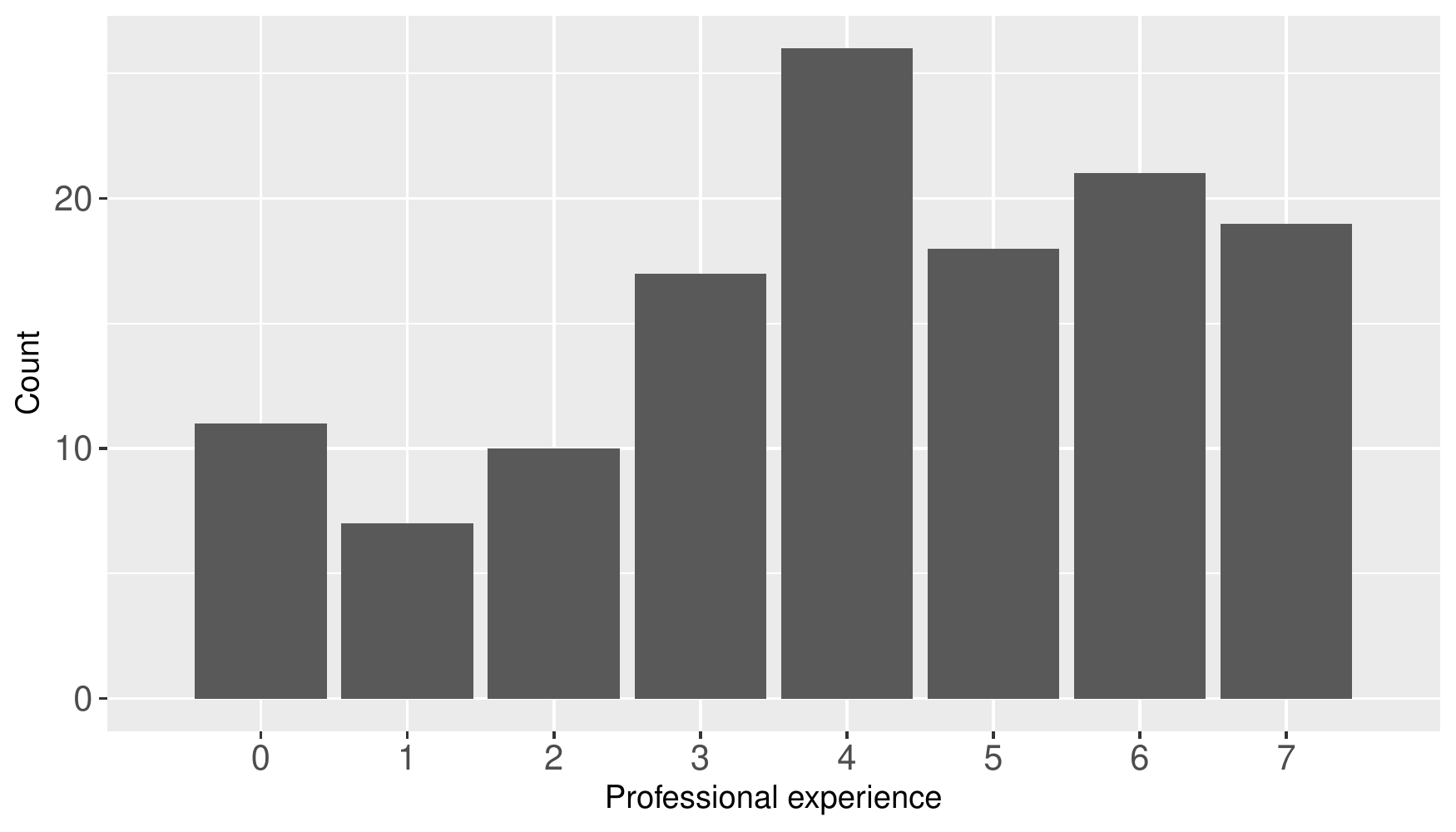}
  \caption{Distribution of professional experience (areas of responsibility).}
  \label{fig:profexp}
\end{figure}

\subsection{Demographics and background variables}

There were 28 (21.7\%) female and 98 (76\%) male respondents; 3 (2.3\%) did not specify gender. Age ranged from 20 to 69 years (MD: 35, SD: 8.7). Company responsibility, highest level of education, and total work experience are shown in Fig.~\ref{fig:demographics}. For company responsibility, we categorized role descriptions into four categories: any kind of developer, as ``Software developer''; any managerial role, from team leader or scrum master to product manager or head of department, as ``Manager''; any kind of analyst or architect role, as ``Analyst / Architect''; and all other roles, including tester or consultant, and unspecified roles, as ``Other / not specified''. For education, a small number of country-specific degrees -- mainly German -- were converted to categories with meaningful similarity.

A measure for \emph{Training} (i.e., academic education or workplace courses) was obtained by asking participants to indicate how much training they had received in each of 12 areas of SE, on a 5-point scale. The sum of the responses was divided by 60 ($5 \times 12$, the maximum score), to yield a score between 0 and 1. This score was then binned into three equal-spaced bins, Low, Medium, and High.

A measure for \emph{Professional experience} was obtained by asking participants to indicate which of seven areas they had been responsible for at some point in their career. The areas were \textit{Requirements}, \textit{Software architecture}, \textit{Software development}, \textit{Software Testing and Quality Assurance}, \textit{Software Configuration Management}, \textit{Project management}, and \textit{Software maintenance}. Participants ticked those in which they had professional experience, yielding an eight-point scale from 0 to 7 by counting the number of selected areas. In addition, participants could indicate other areas, but only six did so, and we did not include those areas in the calculation. The distribution of the professional experience score is shown in Fig.~\ref{fig:profexp}. For analysis, we binned the variable into three equal-spaced bins, Low, Medium, and High.

The level of agility in their working environment as perceived by the participants, was categorized into three levels, low (responses 1-3), medium (response 4), and high (response 5). 


\subsection{Do software professionals exhibit temporal discounting?}

\begin{figure}
  \centering
  \subfloat[Full data for all participants]{\includegraphics[width=0.5\columnwidth]{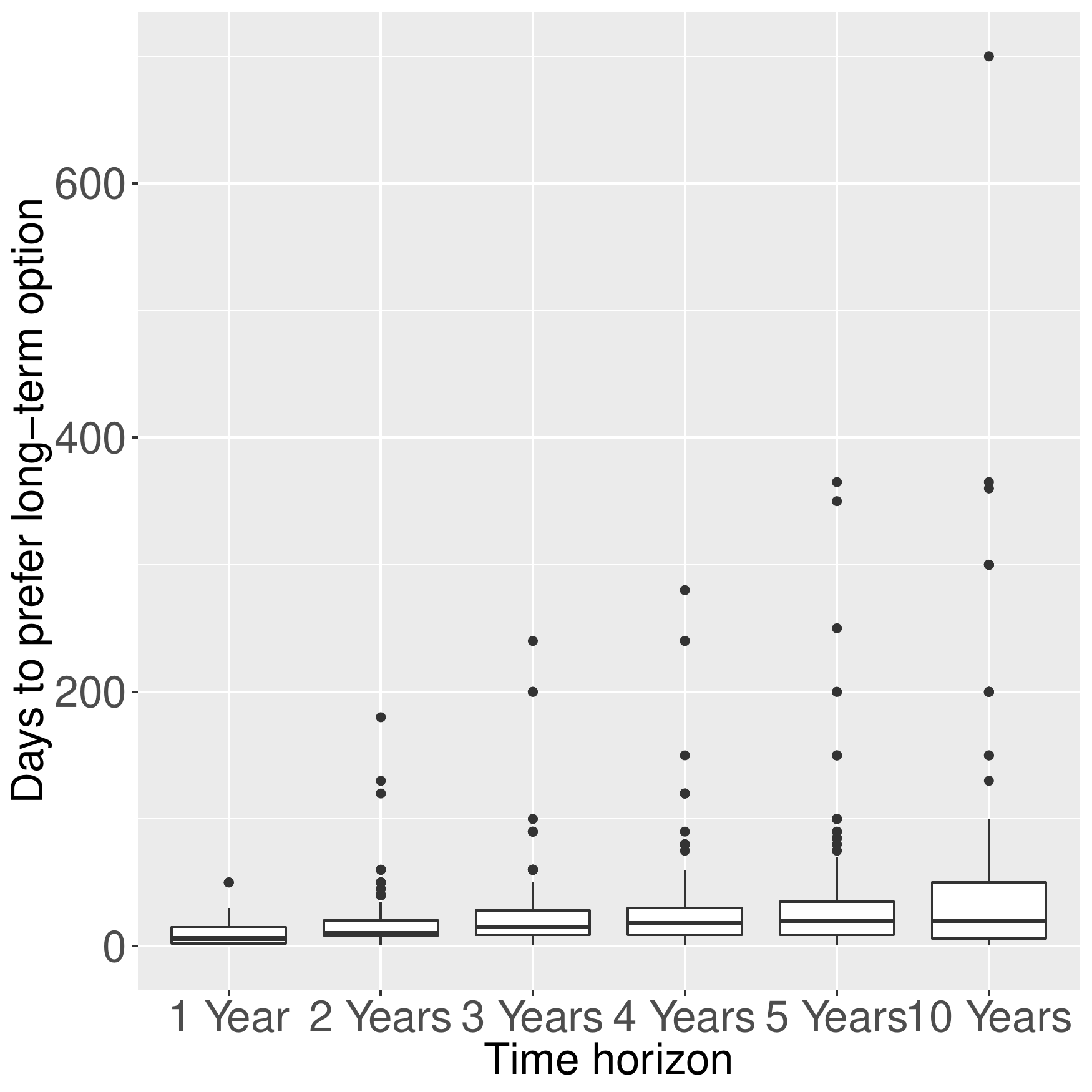}\label{fig:discounting-all-group}}
  \hfill
  \subfloat[Zoomed view of data]{\includegraphics[width=0.5\columnwidth]{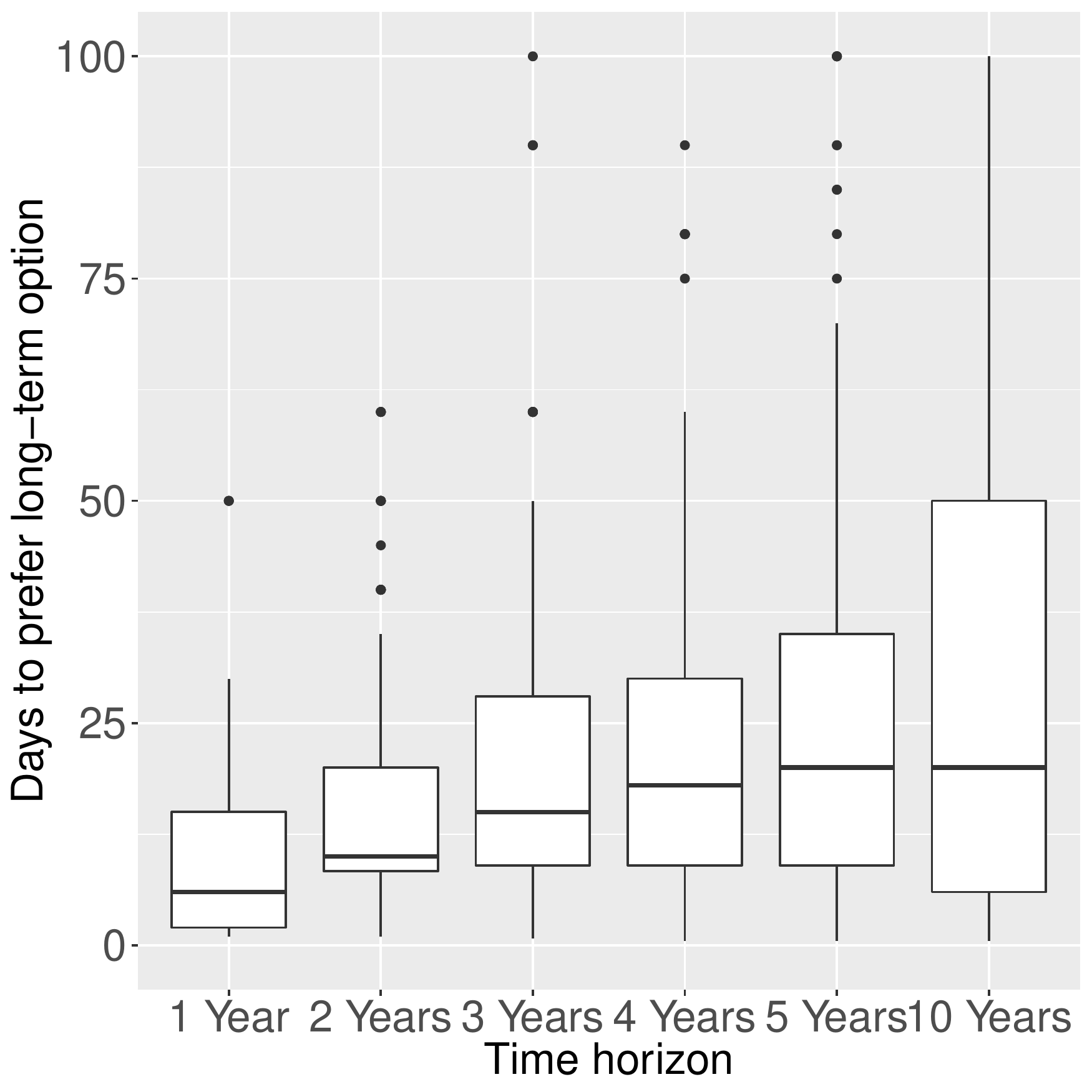}\label{fig:discounting-all-group-zoomed}}
  \caption{Distribution of time savings (days) required to prefer a long-term investment, across different project time horizons. The left figure (a) illustrates the wide variance in discounting. In the zoomed figure (b), outliers above 100 days are omitted to better illustrate the main effect.}
  \label{fig:discounting-all-group-main}
\end{figure}

We first consider the question of temporal discounting across all participants. Fig.~\ref{fig:discounting-all-group-main} depicts the time savings required by the participants to choose the long-term library option for various time horizons. The box plot shows the median number of days for each time horizon (dark line); the 25th and 75th percentiles (bottom and top of the box); minimum and maximum values (horizontal whiskers), and outliers (dots). There is a significant spread of responses, ranging from 1 to 700 (!) days.

\begin{figure}[b]
  \centering
  \includegraphics[width=\columnwidth]{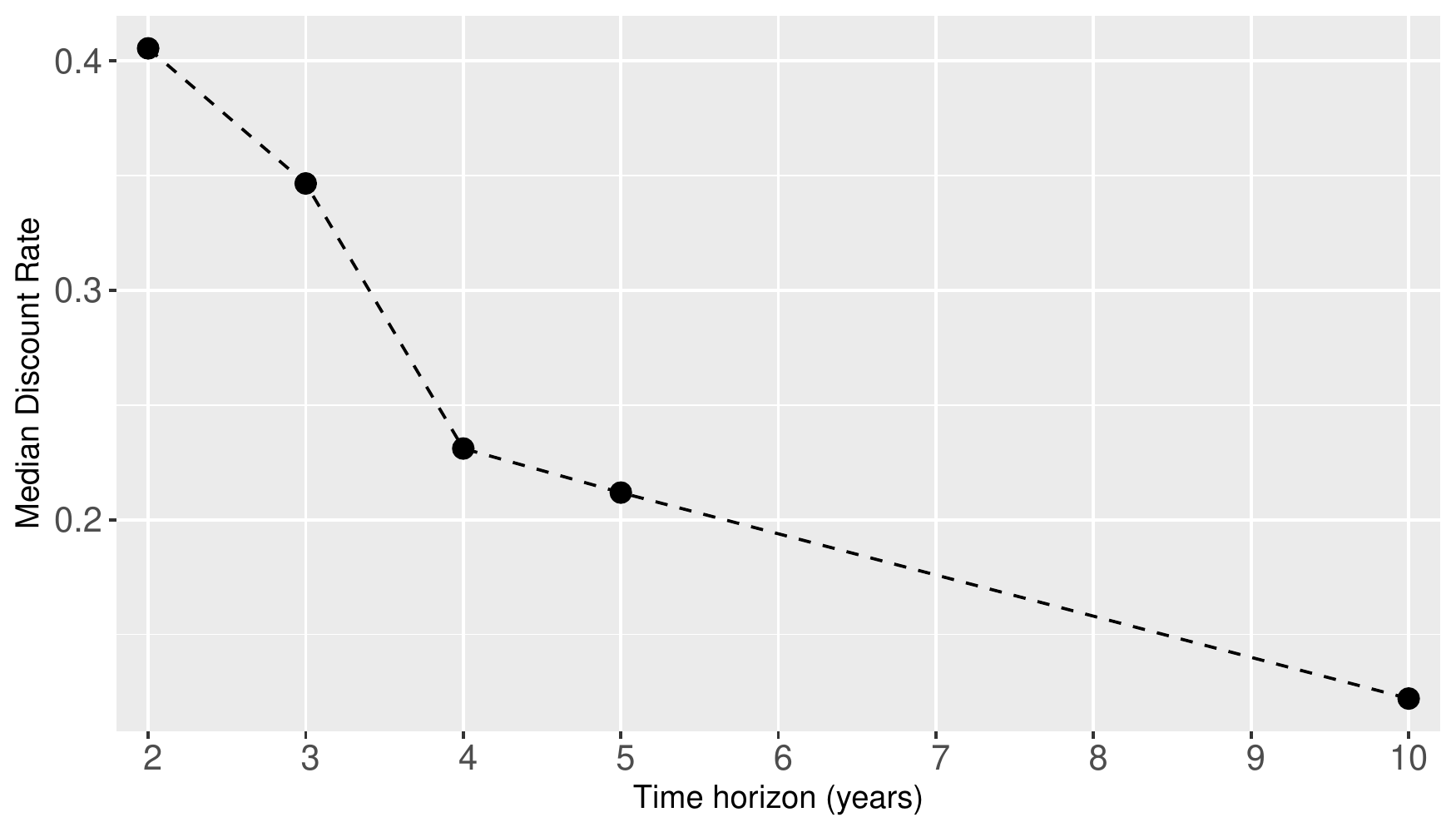}
  \caption{Median discount rate for all participants across project time horizons.}
  \label{fig:discountrate-all}
\end{figure}

Fig.~\ref{fig:discountrate-all} shows the median discount rate of all respondents. The rate is positive and declines over time, indicating that discounting does occur and is most pronounced for early time differences.  We observe large differences among participants, but the overall pattern confirms that distant outcomes are valued lower: in the examined scenario participants demand more savings to opt for the long-term investment. This confirms the results of the original study and other studies on intertemporal choices \cite{frederick2002}, and gives an answer to RQ1.

\subsection{Examining background factors}

We examined the association between AUC and background variables to determine which factors might influence temporal discounting (RQ2 \& RQ3). No association was found for age (using Pearson's correlation test), work experience (using Kendall's rank correlation test), agility, or training (using Kruskal-Wallis test). We did not compare students and professionals due to the highly uneven samples.

\begin{figure}
  \centering
  \includegraphics[width=\columnwidth]{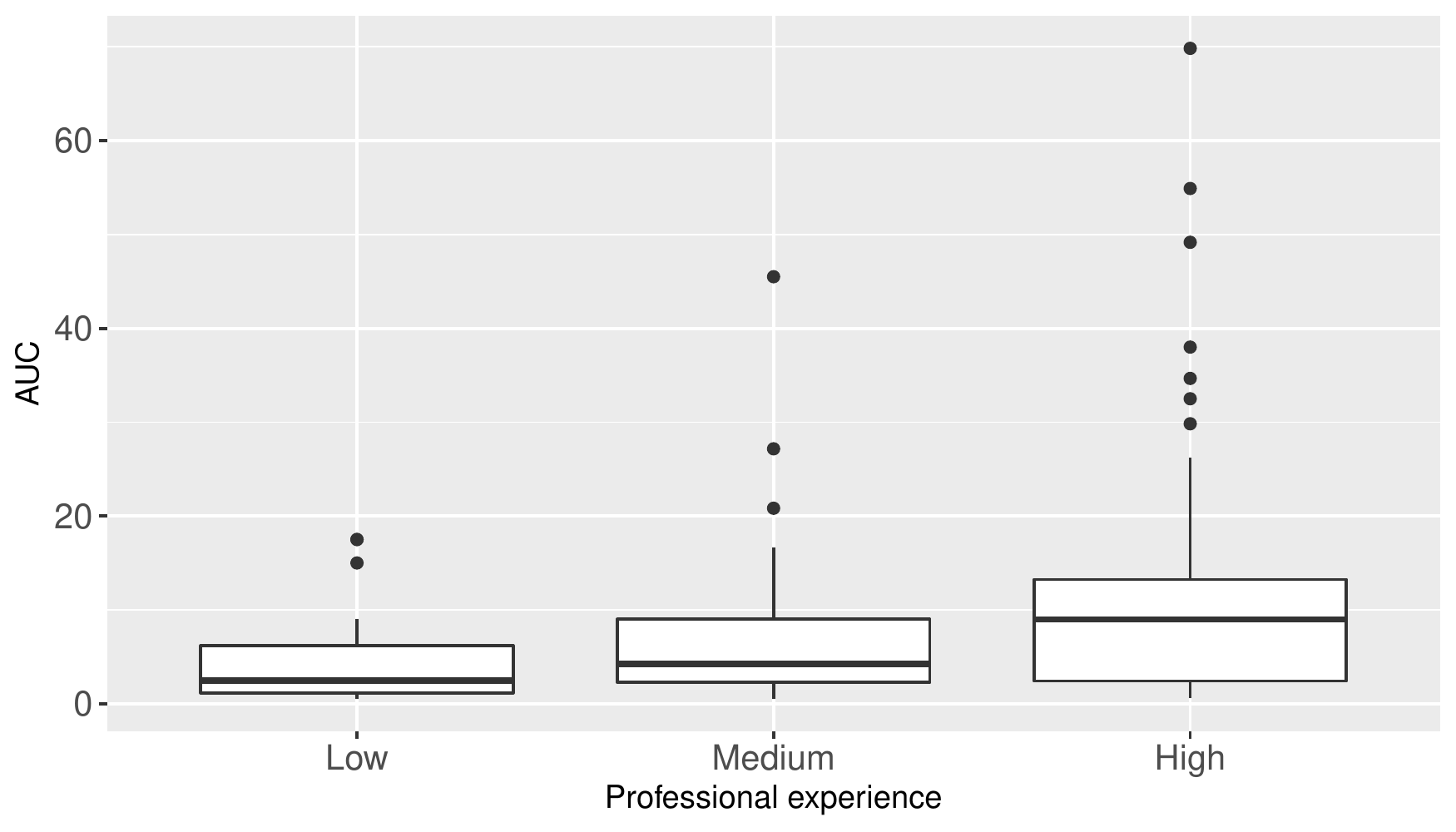}
  \caption{Discounting by professional experience (Medians: low 2.46; medium 4.26; high: 9).}
  \label{fig:association-auc-profexp}
\end{figure}

For professional experience, however, a Kruskal-Wallis test indicated a statistically significant difference at the $p \leq 0.05$ level ($p = 0.0207$) when examining low, medium, and high professional experience. The prerequisite sample distribution shape similarity was met (Anderson-Darling test $p \leq 0.05$). This indicates that the breadth of professional experience to some extent does influence discounting. We used the post-hoc Dunn test to determine which levels of professional experience differ from each other. The Dunn test is appropriate for groups with unequal numbers of observations~\cite{zar2010}. The test indicated that each of the three professional experience conditions is different ($p \leq 0.01$). This is shown in Fig.~\ref{fig:association-auc-profexp}, which shows higher median AUC for higher professional experience, indicating that there is \emph{less} discounting for more breadth in professional experience.

\begin{figure}
      \centering \includegraphics[width=\columnwidth]{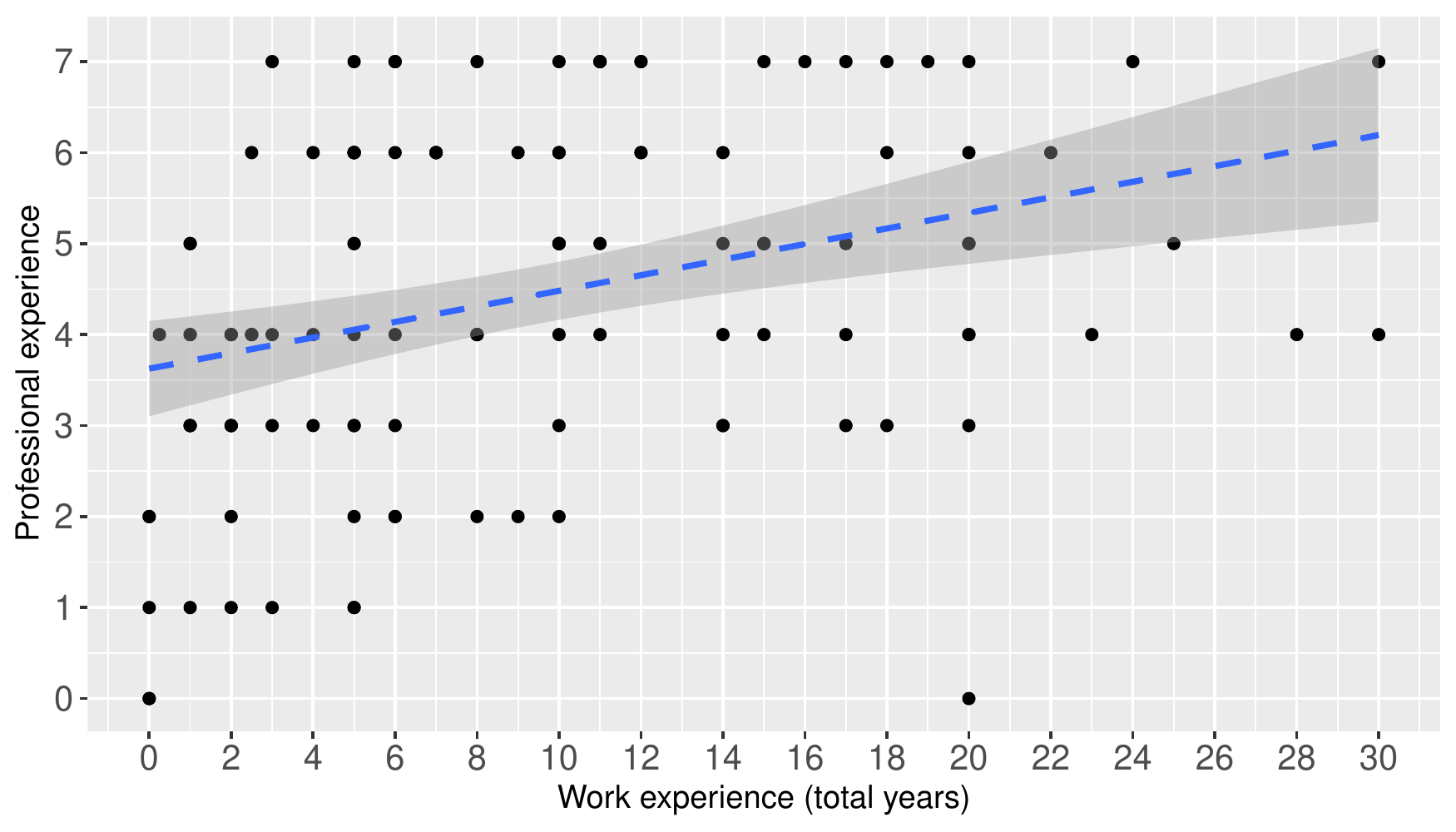}
  \caption{Association between professional experience and length of work experience. The linear regression curve shows a positive relationship. The shaded area represents the confidence interval.}
  \label{fig:association-profexp-workexp}
\end{figure}

The score reflects the breadth of experience rather than length, which was measured by the total work experience variable (in years). The latter was not associated with temporal discounting (AUC). However, since breadth (number of professional areas) and length (years) of experience may be associated, we examined these two variables more closely. A Shapiro-Wilk test indicated that both variables were normally distributed. We therefore tested association using Pearson correlation, since a test for a linear relationship would be more conservative than a rank-order correlation (such as Spearman's $\rho$ or Kendall's $\tau$). We found a weak ($r = 0.337$, 95\% CI: [0.1672, 0.4875]) but significant relationship at the $p \leq 0.01$ level. Examining the relationship graphically (see Fig.\ref{fig:association-profexp-workexp}), we can see that with increased work experience, the lower bound of professional experience does indeed increase. Our interpretation is that increased work experience does increase the likelihood of increased professional experience, but lower work experience does not preclude breadth of professional experience. 

This suggests that it is breadth of experience rather than length that influences temporal discounting (RQ3). This finding is rather striking, but it aligns with prior research in JDM and in SE that shows how temporal and social distance interact in ways that are extremely relevant for central questions raised by this study. The concept of \textit{psychological distance} incorporates temporal and other aspects of distance \cite{fujita_psychology_2015}. The degree to which possible outcomes can be imagined ``viscerally'' greatly influences discounting \cite{weber2006}. This finding from JDM is consistent with a recent study in SE that showed that all else being equal, developers are more likely to recommend that other people's code be fixed than their own \cite{amanatidis_developers_2018}. We speculate that breadth of experience both adds to the cognitive repertoire available to reason heuristically about potential scenarios, as well as increases one's lateral vision and ability to empathize with different positions in projects and different roles, which makes it easier to envision distant outcomes and thus decreases excessive discounting. This is even more interesting when we consider that education in SE areas was \textit{not} associated with any differences in temporal discounting.

We examined the data by country (see Table~\ref{tab:samples}) for differences in discounting (RQ2). However, the difference in AUC variance between countries was too large for a meaningful statistical test. This is expected, since other studies have observed extreme individual variance in discounting~\cite{frederick2002}, which small samples will expose. We observe that countries with at least 10 participants had a median AUC from 2.43 (Germany) to 9 (UK), suggesting that this should be investigated further.

We also examined whether the replicating researcher is associated with discounting. We compared the data collected by one original researcher (sets A-C; $n = 50$) to the data collected by those researchers who were not part of the original study (sets D-L; $n = 79$). We found no significant difference in discounting: a Kruskal-Wallis test indicated no statistically significant difference in AUC for different researchers.

\section{Implications and Threats to validity}

The replication results confirm the occurrence of temporal discounting among software practitioners and students (RQ1). They demonstrate strong variance in discounting between study participants, as found in the original study. Large absolute differences were found between countries, but the general trend was similar, with more discounting for future time horizons (RQ2). The replication contributes new information regarding the influence of background factors on temporal discounting: age, prior education and workplace training, professional experience, and perceived agility of the development team (RQ3). Furthermore, the replication demonstrates the occurrence of discounting in samples of both professional and student participants from different countries and thus cultural backgrounds. The results identify a significant relationship between the breadth of prior professional experience and temporal discounting, which merits further investigation.

The lack of association between education and discounting should raise questions about the effectiveness of current SE teaching in terms of its ability to equip software professionals with means for long-term, sustainable decision-making. It may be that the teaching is not effective at conveying anything that would affect such decision-making, or that the methods and mental models are conveyed but are themselves not adequate. It also points to the long-established sociological finding that plans, and therefore methods, do not determine the course of action but serve merely as a ``weak resource'' in ``situated action''~\cite{suchman_human-machine_2007}. How methods and procedures influence situated decisions needs to be examined using naturalistic decision making research methods \cite{zsambok_naturalistic_1997,klein_decision_1993,klein_naturalistic_2008,zannier_model_2007}. 

The task is arguably simplistic in comparison with the complexities of real-world software development, so any observed effect would not automatically imply similar behavior in a real situation. However, we would expect to see some impact of training on this synthetic scenario. In light of the finding that breadth of experience does influence discounting, we speculate that broadly experienced participants have learned something in their profession that SE teaching, whether academic or in the workplace, does not convey.

\subsection{Threats to validity}

The replication partially mitigated the threats to external validity from which the original study suffered. The fact that temporal discounting has been validated from the responses of 16 participant sets and 129 subjects in total, confirms that software professionals discount future options when confronting a software project management decision with uncertain future outcomes at different points in time. Nevertheless, construct validity threats exist, as the instrument (questionnaire) and the particular scenario cannot shield the subjects' opinion from external effects. It should however be noted that the immunity of temporal discounting to factors such as age, prior education, country of residence, student or professional status, etc., points to a common understanding of the presented task. 

\subsection{Implications for practitioners}

The empirical evidence on the tendency of software professionals to heavily discount distant outcomes underlines the need to think carefully about planning and communication in project management and maintenance tasks with longer-term implications. All else being equal, without further argumentation or incentives, many software developers will generally opt for a small short-term benefit over a significantly larger long-term benefit. This tendency has been decried by many for a long time \cite{neumann_foresight_2012} and is considered by many a root cause of Technical Debt. The current study revealed that developers with broader experience exhibit less discounting; this finding could be taken into account in the assignment of technical tasks related to long-term investments on software quality so as to balance out the dominant tendency of temporal discounting, but also suggests that there are far-reaching indirect consequences of organizational diversity and varied, non-standard career paths. 

\subsection{Implications for researchers}

While it may be tempting to take the quantitative nature of the presented results as a causal explanation, we must guard against premature conclusions on what the findings establish. All we know is that people behave \textit{as if} they would perform temporal discounting. We have not identified how or why this effect takes place, nor do we have a ``gold standard'' of optimal decision making.  There is no optimal decision to be made in the presented scenario, and there are many good reasons for discounting uncertain future outcomes. Many professional situations may be structured in a way that makes temporal discounting perfectly reasonable, be it because of job rotation and turnover, incentive structures, divisions of labour, business models, project cycles, or other factors.

Deviations from supposed normative ideals of ``rational'' decision making are often prematurely labelled as ``human error'' and ``cognitive bias'', but JDM researchers have for decades demonstrated the value of the alternative reading: If empirical findings contradict normative models, this often points to misguided assumptions in the normative models~\cite{beach_why_1993}. In intertemporal choice, for example, an experiment established that an alternative and very robust explanation for temporal discounting lies not in time-based discount factors, but differences in subjective time \textit{perception}~ \cite{zauberman_discounting_2009}. ``Naturalistic'' JDM research, as opposed to ``rationalistic'' research, focuses on descriptive rather than normative methods and models, with interesting and highly relevant implications~\cite{zsambok_naturalistic_1997,klein_naturalistic_2008,bell_decision_1989,kahneman_conditions_2009,zannier_model_2007,becker2017}. This tension between normative and descriptive research is also reflected in recent discussions in SE \cite{ralph_two_2018}.

The effect of temporal discounting is established. We must now examine its patterns, mechanisms, factors, effects, and possible interventions. A host of questions await examinations, and a set of conceptual frameworks from JDM can be brought to bear on them. Which factors affect discounting most? In what patterns does temporal discounting occur in SE, and where are its effects most pronounced? How does the differential discounting of gains, losses and mixed outcomes~\cite{soman2005} manifest in SE? How can we reduce excessive discounting in specific areas such as project management or technical debt? Which assumptions of current methods in these areas need to be revisited~\cite{becker2017}? How should the findings influence the creation of new methods for use in industry? How should educators prepare future generations of software professionals to sustainably construct and maintain software systems with increasing complexity and impact on business and society? Combining future conceptual replications with the methods of Cognitive Task Analysis~\cite{crandall_working_2006} will be essential to construct a richer understanding of the macro-cognitve landscape of SE practice.

\section{Conclusions}

Temporal discounting research studies the relative valuation placed on foreseeable benefits at different points in time, the mechanisms by which individuals and groups establish their preferences, and the choices that result from this. Evidence across a number of fields shows that proximal rewards are weighted higher than distant ones. 

A recent questionnaire-based study investigated the existence of temporal discounting by software professionals. We have replicated that study in 16 different populations. The results validate that software professionals exhibit temporal discounting. Moreover, we examined the association between background variables and the Area Under Curve (AUC) as an aggregate measure of discount rate. The statistical analysis indicates that temporal discounting is not influenced by factors such as age, length of work experience, amount of training, or the level of agility in participants' environments. However, breadth of professional experience was found to influence discounting: participants with broader professional experience exhibit less discounting. This is a striking finding that ties in with prior research from JDM and opens new avenues of research. Software engineering practice and software maintenance in particular regularly face situations where long-term options must be weighed against short-term ones. Drawing robust methods from Judgement and Decision Making and building a solid empirical grounding can substantially improve our knowledge on the factors that influence software professionals when making intertemporal decisions.  

\section*{Acknowledgments}

CB suggested the initial idea. Further ideation and study design was performed jointly by FF, CB, AC and RM. AC, SB, LD, BP and CV collected the data for the study. FF coordinated the data collection. FF and AC performed most of the data analysis. FF, CB and AC led the writing. All authors provided comments and approved the final text.

This research was partially supported by the Natural Sciences and Engineering Research Council through RGPIN-2016-06640, the Canada Foundation for Innovation, the Ontario Research Fund, and the KKS Foundation through the S.E.R.T. Research Profile at Blekinge Institute of Technology. The research leading to these results has received funding from the European Union's Horizon 2020 research and innovation programme under the Marie Sk\l{}odowska-Curie grant agreement No 712949 (TECNIOspring PLUS) and from the Agency for Business Competitiveness of the Government of Catalonia.

\IEEEtriggeratref{45}
\bibliographystyle{IEEEtran}
\bibliography{citations}

\end{document}

%% file: tables/survey-demographics.tex
\begin{tabular}{p{.4\columnwidth} p{.5\columnwidth}}
\toprule
Item & Description \\
\midrule
Gender & Female / Male / Other \\
Year of Birth & Numeric input \\
\addlinespace
\midrule
\multicolumn{2}{l}{\textit{Educational background}} \\
Highest completed degree & Bachelor / Masters / Doctorate / Other \\
Field of degree & Computer Science / Other \\
Training in 12 SWEBOK areas & A 5-point scale ranging from ``None or almost none'' to ``A lot'' for each area \\
\addlinespace
\midrule
\multicolumn{2}{l}{\textit{Professional background}} \\
Current company role & Free text input \\
Professional experience in 7 areas of software development & A true/false choice indicating experience in each area \\
Perceived team agility & A 5-point scale ranging from ``very plan-driven'' to ``very agile'' \\
Work experience & Numeric input for total work experience, total in current company, and total in current role \\
\bottomrule
\end{tabular}

%% file: tables/descriptive-samplesizes.tex
\begin{tabular}{lllrr}
\toprule
Set & Country & Type & Size & Running tot.\\
\midrule
A & Switzerland & Prof. (same company) & 3 & 3\\
B & Romania & Prof. (same company) & 3 & 6\\
C & Greece & Prof. (same company) & 44 & 50\\
D & Germany & Various professionals & 16 & 66\\
E & UK & Prof. (same company) & 11 & 77\\
\addlinespace
F & UK & Prof. (same company) & 1 & 78\\
G & UK & Prof. (same company) & 3 & 81\\
H & UK & Prof. (same company) & 1 & 82\\
I & UK & Prof. (same research org.) & 20 & 102\\
J & UK & Students & 2 & 104\\
\addlinespace
K & Brazil & Prof. (same company) & 2 & 106\\
L & Brazil & Prof. (same company) & 4 & 110\\
M & Brazil & Various professionals & 6 & 116\\
N & Germany & Prof. (same company) & 1 & 117\\
O & Germany & Prof. (same company) & 1 & 118\\
\addlinespace
P & Germany & Students & 11 & 129\\
\bottomrule
\end{tabular}